\documentclass[twocolumn,floatfix]{aastex631}

\usepackage[caption=false]{subfig}

\begin{document}

\title{Small-scale cosmic ray anisotropy observed by the GRAPES-3 experiment at TeV energies}

\author{M.~Chakraborty}   \affiliation{Tata Institute of Fundamental Research, Homi Bhabha Road, Mumbai 400005, India}
\author   {S.~Ahmad}         \affiliation{Aligarh Muslim University, Aligarh 202002, India}

\author   {A.~Chandra}       \affiliation{Aligarh Muslim University, Aligarh 202002, India}
\author   {S.R.~Dugad}       \affiliation{Tata Institute of Fundamental Research, Homi Bhabha Road, Mumbai 400005, India}
\author   {U.D.~Goswami}     \affiliation{Dibrugarh University, Dibrugarh 786004, India}
\author   {S.K.~Gupta}       \affiliation{Tata Institute of Fundamental Research, Homi Bhabha Road, Mumbai 400005, India}
\author   {B.~Hariharan}     \affiliation{Tata Institute of Fundamental Research, Homi Bhabha Road, Mumbai 400005, India}
\author   {Y.~Hayashi}      \affiliation{Graduate School of Science, Osaka Metropolitan University, Sugimoto, Sumiyoshi, Osaka 558-8585, Japan}
\author   {P.~Jagadeesan}    \affiliation{Tata Institute of Fundamental Research, Homi Bhabha Road, Mumbai 400005, India}
\author   {A.~Jain}          \affiliation{Tata Institute of Fundamental Research, Homi Bhabha Road, Mumbai 400005, India}
\author   {P.~Jain}          \affiliation{Indian Institute of Technology Kanpur, Kanpur 208016, India}
\author   {S.~Kawakami}      \affiliation{Graduate School of Science, Osaka Metropolitan University, Sugimoto, Sumiyoshi, Osaka 558-8585, Japan}
\author   {T.~Koi}           \affiliation{College of Engineering, Chubu University, Kasugai, Aichi 487-8501, Japan}
\author   {H.~Kojima}        \affiliation{Chubu Astronomical Observatory, Chubu University, Kasugai, Aichi 487-8501, Japan}
\author   {S.~Mahapatra}     \affiliation{Utkal University, Bhubaneshwar 751004, India}
\author   {P.K.~Mohanty}
%\email{pkm@tifr.res.in}
\affiliation{Tata Institute of Fundamental Research, Homi Bhabha Road, Mumbai 400005, India}
\author   {R.~Moharana}      \affiliation{Indian Institute of Technology Jodhpur, Jodhpur 342037, India}
\author   {Y.~Muraki}        \affiliation{Institute for Space-Earth Environmental Research, Nagoya University, Nagoya 464-8601, Japan}
\author   {P.K.~Nayak}       \affiliation{Tata Institute of Fundamental Research, Homi Bhabha Road, Mumbai 400005, India}
\author   {T.~Nonaka}        \affiliation{Institute for Cosmic Ray Research, Tokyo University, Kashiwa, Chiba 277-8582, Japan}
\author   {T. Nakamura}      \affiliation{Faculty of Science and Technology, Kochi University, Kochi 780-8520, Japan}
\author   {A.~Oshima}        \affiliation{College of Engineering, Chubu University, Kasugai, Aichi 487-8501, Japan}
\author   {B.P.~Pant}        \affiliation{Indian Institute of Technology Jodhpur, Jodhpur 342037, India}
\author   {D.~Pattanaik}     \affiliation{Tata Institute of Fundamental Research, Homi Bhabha Road, Mumbai 400005, India} \affiliation{Utkal University, Bhubaneswar 751004, India}
\author   {S.~Paul}          \affiliation{Tata Institute of Fundamental Research, Homi Bhabha Road, Mumbai 400005, India}
\author   {G.S.~Pradhan}     \affiliation{Indian Institute of Technology Indore, Indore 453552, India}
\author   {M.~Rameez}        \affiliation{Tata Institute of Fundamental Research, Homi Bhabha Road, Mumbai 400005, India}
\author   {K.~Ramesh}        \affiliation{Tata Institute of Fundamental Research, Homi Bhabha Road, Mumbai 400005, India}
\author   {S.~Saha}          \affiliation{Indian Institute of Technology Kanpur, Kanpur 208016, India}
\author   {R.~Sahoo}         \affiliation{Indian Institute of Technology Indore, Indore 453552, India}
\author   {R.~Scaria}        \affiliation{Indian Institute of Technology Indore, Indore 453552, India}
\author   {S.~Shibata}       \affiliation{College of Engineering, Chubu University, Kasugai, Aichi 487-8501, Japan}
\author   {T.~Tabata}         \affiliation{College of Engineering, Chubu University, Kasugai, Aichi 487-8501, Japan}
\author   {H.~Takamaru}       \affiliation{College of Engineering, Chubu University, Kasugai, Aichi 487-8501, Japan}
\author  {K.~Tanaka}           \affiliation{Graduate School of Information Sciences, Hiroshima City University, Hiroshima 731-3194, Japan}
\author   {F. Varsi}         \affiliation{Indian Institute of Technology Kanpur, Kanpur 208016, India}
\author   {K.~Yamazaki}      \affiliation{College of Engineering, Chubu University, Kasugai, Aichi 487-8501, Japan}
\author   {M.~Zuberi}        \affiliation{Tata Institute of Fundamental Research, Homi Bhabha Road, Mumbai 400005, India}

\correspondingauthor{P.K. Mohanty}
\email{pkm@tifr.res.in}

%\collaboration{20}{(The GRAPES-3 Collaboration)}



\begin{abstract}

GRAPES-3 is a mid-altitude (2200 m) and near equatorial ($11.4^{\circ}$ North) air shower array, overlapping in its field of view for  cosmic ray observations with experiments that are located in Northern and Southern hemispheres. We analyze a sample of $3.7\times10^9$ cosmic ray events collected by the GRAPES-3 experiment between 1 January 2013 and 31 December 2016 with a median energy of $\sim16$ TeV for study of small-scale ($<60^{\circ}$) angular scale anisotropies. We observed two structures labeled as A and B, deviate from the expected isotropic distribution of cosmic rays in a statistically significant manner. Structure `A' spans $50^{\circ}$ to $80^{\circ}$ in the right ascension and $-15^{\circ}$ to $30^{\circ}$ in the declination coordinate. The relative excess observed in the structure A is at the level of $(6.5\pm1.3)\times10^{-4}$ with a statistical significance of 6.8 standard deviations. Structure `B' is observed in the right ascension range of $110^{\circ}$ to $140^{\circ}$. The relative excess observed in this region is at the level of $(4.9\pm1.4)\times10^{-4}$ with a statistical significance of 4.7 standard deviations. These structures are consistent with those reported by Milagro, ARGO-YBJ, and HAWC. These observations could provide a better understanding of the cosmic ray sources, propagation and the magnetic structures in our Galaxy.

\end{abstract}



\section{Introduction} \label{sec:intro}
Cosmic rays (CRs) are charged particles because of which they undergo deflections by the randomized magnetic field in the interstellar medium through which they propagate. This leads to an isotropic distribution of CRs as observed on the Earth. However, in the past decade, several experiments located at different latitudes have observed the CRs anisotropy on different angular scales and energy ranges with an amplitude of $\sim\!10^{-4}$ to $10^{-3}$. Large-scale structures with a dominant dipolar component have been observed in TeV--PeV energy range by several ground based extensive air shower (EAS) experiments located in the Northern hemisphere such as Tibet-AS$\gamma$ \citep{Amenomori2006, Amenomori2017}, ARGO-YBJ \citep{Bartoli2018}, Milagro \citep{Abdo2009}, HAWC \citep{Abeysekara2019}, EAS-TOP \citep{Aglietta2009}, KASCADE-Grande \citep{Ahlers2019} and in Southern hemisphere such as IceCube \citep{Aartesen2016}. An excess has been observed within the right ascension ($\alpha$) of $30^{\circ}$ to $120^{\circ}$ and a large deficit region has been observed within $150^{\circ}\leq\alpha\leq250^{\circ}$. Anisotropies at ultra high energies ($\ge\! 10^{18}$ eV) where the interstellar magnetic field plays a relatively insignificant role for bending CRs, has been reported by the Pierre Auger Observatory \citep{Aab2018} and the Telescope Array \citep{Abbasi2020}. The large-scale anisotropy with a strength of $\sim\!10^{-3}$ has been qualitatively explained by standard diffusion models describing propagation of CRs through the Galaxy following acceleration by supernova remnants \citep{Blasi2012, Giacinti2012, Mertsch2015}. 

Small-scale anisotropy with angular width less than $60^{\circ}$ was first reported by the Milagro experiment showing excesses in two regions, namely region A and B with an excess at the level of $6.0\times10^{-4}$ and $4.0\times10^{-4}$ respectively \citep{Abdo2008}. Located at $36^{\circ}$N latitude, the Milagro could observe the upper half of region A centered at $\alpha\approx69.4^{\circ}$, $\delta\approx13.8^{\circ}$, ranging in $66^{\circ}<\alpha < 76^{\circ}$ and $10^{\circ}<\delta <20^{\circ}$. The region B was reported to be a structure lying within $15^{\circ}< \delta < 50^{\circ}$ with right ascension within $117^{\circ}< \alpha < 141^{\circ}$. The ARGO-YBJ also observed the same regions along with two additional regions namely `3' and `4' \citep{Bartoli2013}. The region 3 stretches in the Northern hemisphere within $234^{\circ}\leq \alpha \leq 282^{\circ}$. The region `4' is observed with $200^{\circ}\leq \alpha \leq 216^{\circ}$ and $24^{\circ}\leq \delta \leq 34^{\circ}$ and is similar to the the region `C' observed by HAWC \citep{Abeysekara2014}. IceCube has also observed small-scale anisotropic structures in the Southern hemisphere by subtracting dipole and quadrapolar terms from the large-scale maps, and the region B appears to have a continuity in the Southern hemisphere \citep{Aartesen2016}. A full sky analysis of HAWC and IceCube combined also shows the two primary anisotropy structures in region `A' and `B' in which the region `B' can be seen to extend throughout the entire declination range \citep{Abeysekara2019}. GRAPES-3 covers a declination range from $-23.8^{\circ}\leq \delta \leq 46.6^{\circ}$ with a 56\% sky coverage and an overlapping field of view with the above discussed experiments. This paper reports the small-scale anisotropy structures and their features observed by the GRAPES-3 experiment.

\section{The GRAPES-3 experiment and data set}

\begin{figure}[htb]
    \includegraphics[width=\linewidth]{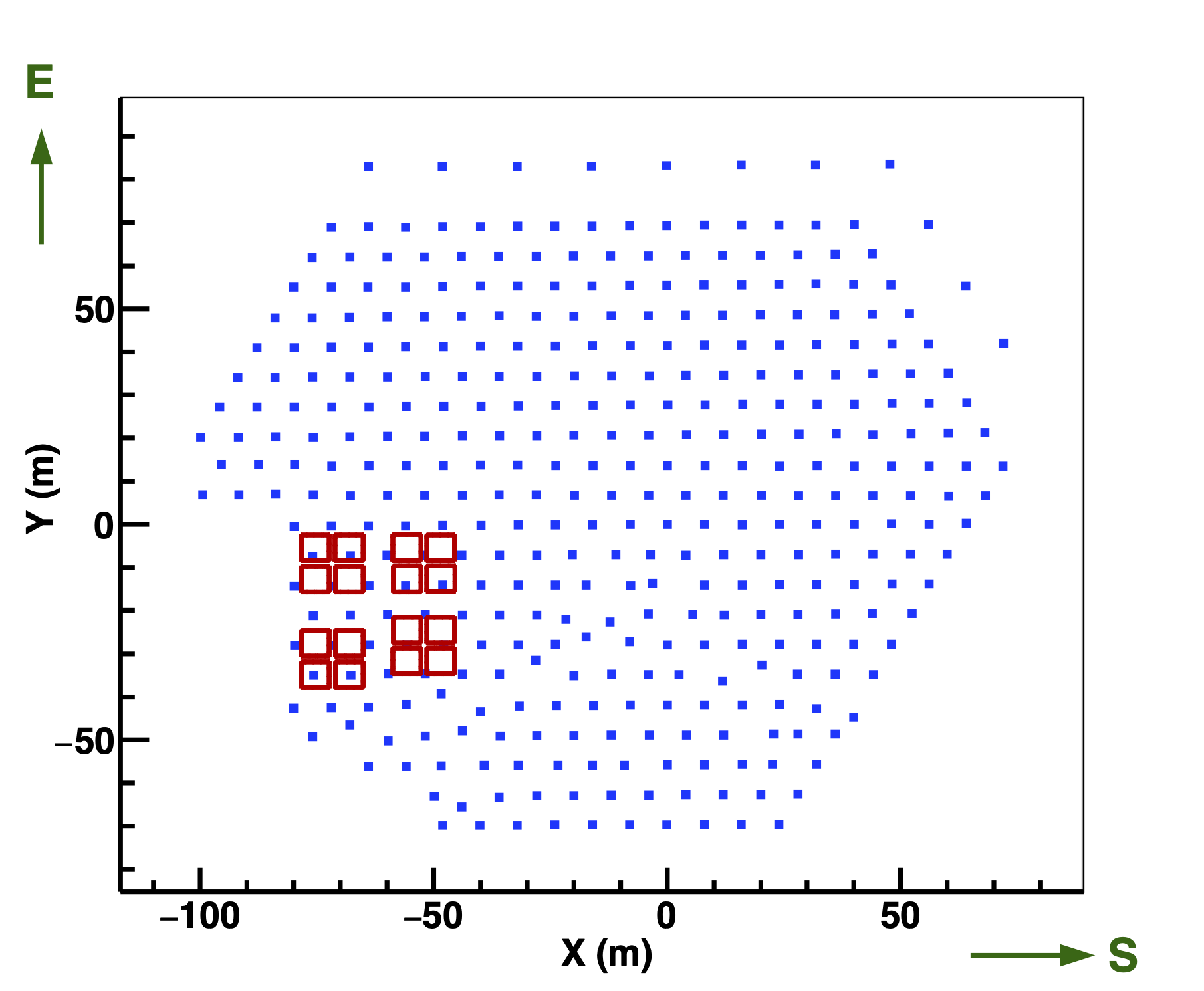}
    \caption{{Schematic diagram of the GRAPES-3 experiment with the scintillator detectors represented by (\textcolor{blue}{$ \blacksquare $}) and the muon detector modules represented by (\textcolor{red}{$\square$}).}}
    \label{fig:fig1}
\end{figure}

The GRAPES-3 is a near equatorial and mid-altitude EAS experiment, located in Ooty, India (11.4$^\circ$N, 76.7$^\circ$E and 2200 m a.s.l.), designed for cosmic ray and gamma ray observations at TeV-PeV energies. It consists of an array of 400 plastic scintillator detectors spread over 25,000\,m$^2$ area \citep{Gupta2005, Mohanty2009} and a tracking muon detector of 560 m$^2$ area \citep{Hayashi2005} as shown schematically in \autoref{fig:fig1}. The scintillator detectors are of 1\,m$^{2}$ area each, placed in a hexagonal configuration with 8\,m inter-detector separation. The density of secondary particles and their relative arrival times in each EAS are recorded by the scintillator detectors. The EAS parameters such as core location, size and age were obtained by fitting Nishimura-Kamata-Greisen (NKG) function to the lateral distribution of particle densities \citep{kamata1958, Greisen1960}. The direction of individual EAS in terms of zenith and azimuth angle ($\theta$, $\phi$) were obtained by a fitting a plane front to the relative arrival time data recorded on each scintillator detector after correcting the curvature in the shower front which is observed to be dependent on both size and age parameter \citep{Jhansi2020, Pattanaik2022}. About 20$\%$ of the EAS events mainly of low energies could not be reconstructed properly by the NKG function. Hence, the curvature corrections could not be performed for these events as the size and age parameters are not reliable. However,  the $\theta$ and $\phi$ for these events were obtained only using a plane fit. This is not expected to impact the study of anisotropy since the angular scale of anisotropies relevant to this analysis are much larger than the improvement in resolution due to curvature corrections. For this analysis, we used 4 years of EAS data recorded from 1 January 2013 to 31 December 2016 which comprised $3.9\times10^9$ EAS events with a recorded live time of 1273.1 days. The EASs with zenith angle less than $60^{\circ}$ were selected. A total of $3.7\times10^9$ EASs which passed the selection criteria were used for the analysis. The median energy of the EAS was calculated to be 16.2 TeV based on energy-size relation using CORSIKA simulation as described in \citep{Pattanaik2022}.

\section{Analysis and results} \label{sec:meth}

In this analysis, the reconstructed local direction ($\theta$, $\phi$) of each EAS event was converted into equatorial coordinate in terms of $\alpha$ and $\delta$. The events were binned into a map in the celestial sphere using HEALPix package \citep{Gorski2005}. We set the parameter nside=64 which divides the sky into 45192 pixels of equal area of about $(0.92^{\circ})^2$ in $\alpha$ and $\delta$. The EAS trigger rate exhibits variations caused by atmospheric pressure and temperature on a daily time scale at the level of 3-5$\%$ \citep{Meeran2017}. Further, there are gaps in the data produced either due to scheduled maintenance activity or failure of electronics or DAQ system. Fluctuation in the rates due to abnormality in the detectors or electronics is another effect. It is difficult to identify and remove all these effects completely from the data. They can cause non-uniform exposure of events in the Celestial sphere, thereby producing spurious anisotropies which could be more than a order of magnitude larger than the  genuine anisotropy of astrophysical origin that needs to be investigated. Therefore, the search for anisotropies in this map cannot be performed directly. Our search strategy is based on the estimation of a reference map which retains the anisotropies originated from instrumental or atmospheric effects while being insensitive to the astrophysical anisotropy. The reference map was estimated from the data itself using a method called time-scrambling as described in \citep{Alexandreas1993}. This method has been employed by several air shower experiments for successful extraction of anisotropy \citep{Bartoli2013,Aartesen2016}. In this method, the true recorded time of an EAS event is forgotten while assigning another time to it which is randomly picked up from a sample of recorded events within a time window of $\Delta t$. The ($\theta$, $\phi$) for the event is kept same. Thus it changes the $\alpha$ of the event while keeping $\delta$ unchanged. Thus, the anisotropic structures of astrophysical origin smaller than an angular scale of $\Delta t \times 15^{\circ}/ 1\,$hr are removed. Each recorded event is assigned with 20 random time values from the recorded time sample within the scrambling time window, and then it is filled into the reference map ($N_i^{ref'}$). This essentially means that the reference map has 20 times more number of events than the data map. This is done to reduce the statistical fluctuation in the reference map. An average of all these 20 maps is then calculated to estimate the reference map as, $N^{ref}_i = N_i^{ref'}/20$. To re-state it, both the reference map and data map will contain the anisotropies produced by atmospheric and instrumental effects whereas the reference map will not have the anisotropies of astrophysical origin. The relative intensity for each pixel is calculated using,
$\delta I= \frac{N^{data}_{i} - N^{ref}_{i}}{N^{ref}_{i}}$
where $N^{data}_{i}$ and $N^{ref}_{i}$ are the number of events in the i-th pixel of data and reference map respectively. 

\begin{figure}[htb!]
  \centering
   \subfloat{\includegraphics[scale=0.33]{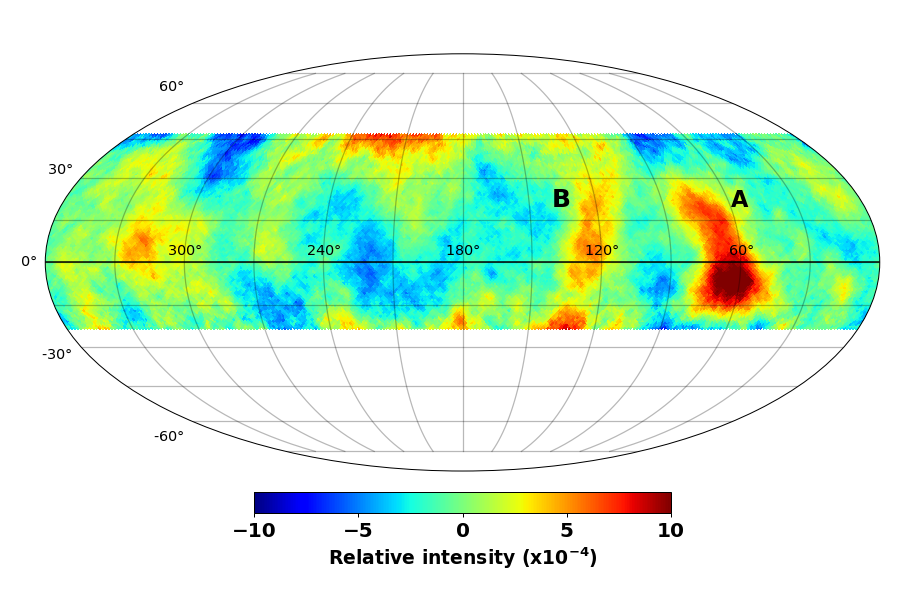}}
  \qquad
  \subfloat{\includegraphics[scale=0.33]{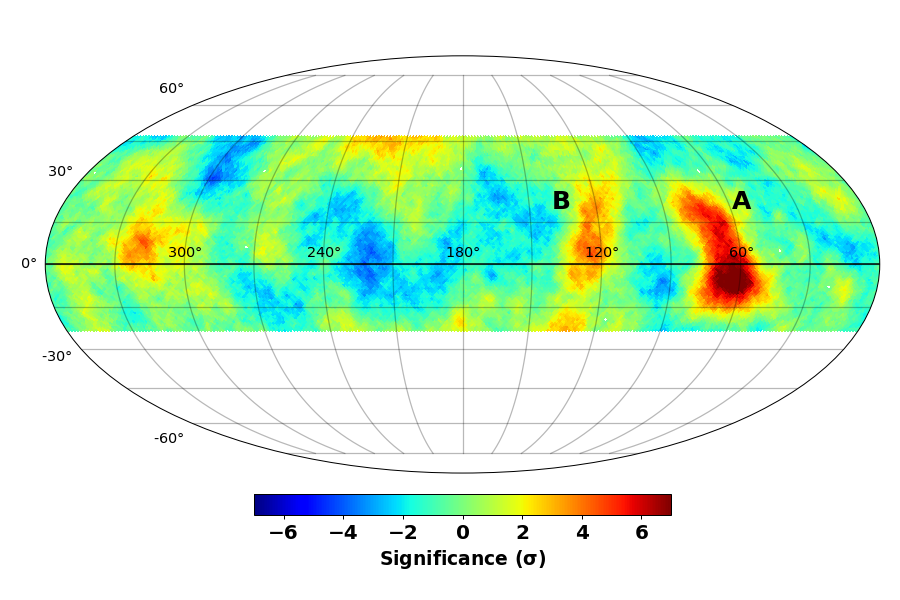}}
 \caption{Anisotropy (top) and significance (bottom) obtained after time-scrambling with $\Delta t=24\,$ hrs and smoothing radius of $10^{\circ}$. The large-scale deficit can be seen while small scale structures A and B are prominently seen.}
  \label{fig:fig2}
\end{figure}

\begin{figure}[ht]
  \centering
   \subfloat{\includegraphics[scale=0.33]{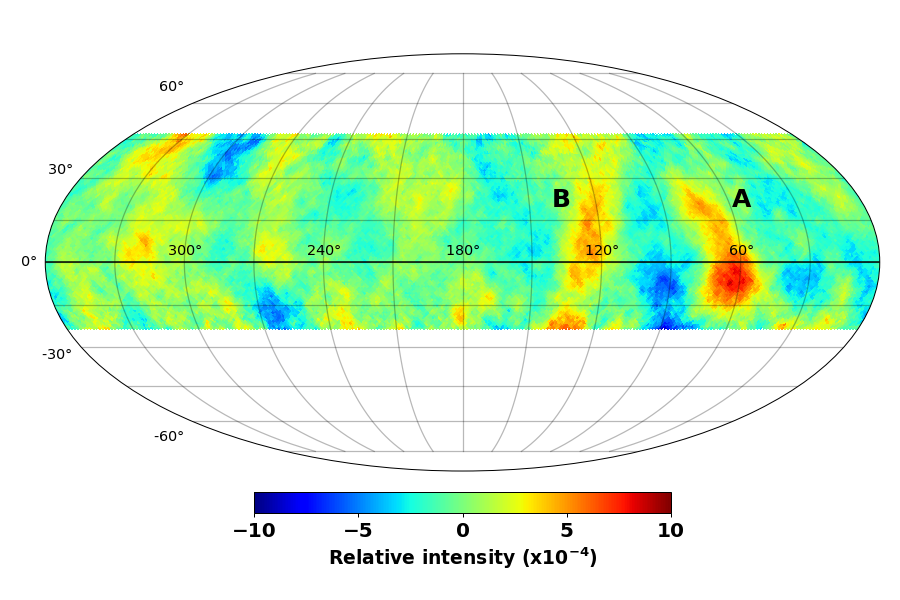}}
   \qquad
  \subfloat{\includegraphics[scale=0.33]{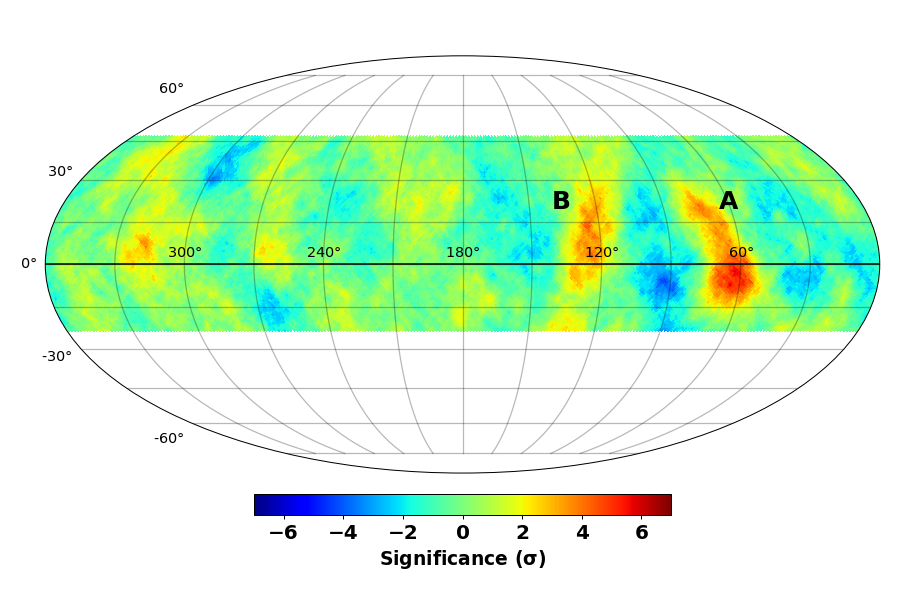}} 
 \caption{Anisotropy (top) and significance (bottom) observed with a scrambling window of 4 hrs and a smoothing radius of $10^{\circ}$}
  \label{fig:fig3}
\end{figure}

First we apply the time-scrambling method on data using a scrambling window of $\Delta t = 24\,$hrs. The stability in zenith and azimuth over long periods of time allows such a choice. Since in 24 hrs, there will be full coverage of the celestial sphere, anisotropies up to the largest angular scale present in the data should show up as they will be destroyed in the reference map. The relative intensity map does not show any discernible anisotropy due to high level of statistical fluctuation in the pixels. To improve the capability of identifying directional features, we employ a smoothing method similar to the method followed by \citep{Abeysekara2014}. This approach includes combining the event counts within individual pixels and adding counts from neighboring pixels positioned within a chosen smoothing radius of $10^\circ$ in this particular study. This approach serves to decrease statistical fluctuations in nearby pixels, thereby revealing any localized surpluses. However, it also establishes correlations in event statistics among neighboring pixels. Essentially, this is analogous to applying a smoothing operation to the map using a top-hat function of $10^\circ$ radius. The structures that we are trying to observe span over few tens of degrees, so a choice of a $10^{\circ}$ smoothing radius has ensured that these structures are not integrated over. The relative intensity map after smoothing is shown in \autoref{fig:fig2}. The pixel-wise significance was calculated using the Li-Ma prescription \citep{LiMa} and the significance map is shown on the bottom of the same figure. The map shows a large-scale deficit region in the right ascension range of 135$^{\circ}$ to 300$^{\circ}$. The location is consistent with the observation by Tibet-AS$\gamma$ \citep{Amenomori2017}, HAWC \citep{Abeysekara2018} and IceCube \citep{Aartesen2016}. Two excess regions with angular scales $\leq$60$^{\circ}$ marked as A and B are more prominent than the large-scale deficit region. The observed strength of the large-scale deficit region is significantly less than expected. This is anticipated due to the attenuation of the reconstructed anisotropy by mid- or low-latitude detectors whose instantaneous field of view is much smaller than the size of the large-scale anisotropy. There are methods developed to retrieve them \citep{Ahlers2016}. However, the current work has focused on the study of small scale anisotropies. Since the structures are of angular size of $\leq$60$^{\circ}$ as can be seen from \autoref{fig:fig2}, we have chosen a time-scrambling window of 4 hrs that can destroy anisotropy up to angular scale of 60$^{\circ}$ in the reference map while retaining the large scale anisotropies. In other words, the relative intensity map will be devoid of the large-scale anisotropies while being sensitive to the small-scale anisotropy structures as shown in \autoref{fig:fig3}. The two structures A and B are largely consistent with the observation of other experiments such as Milagro \citep{Abdo2008}, HAWC \citep{Abeysekara2014} and ARGO-YBJ \citep{Bartoli2013}. 

\begin{figure}[!t]
    \centering
    \includegraphics[width=1\linewidth]{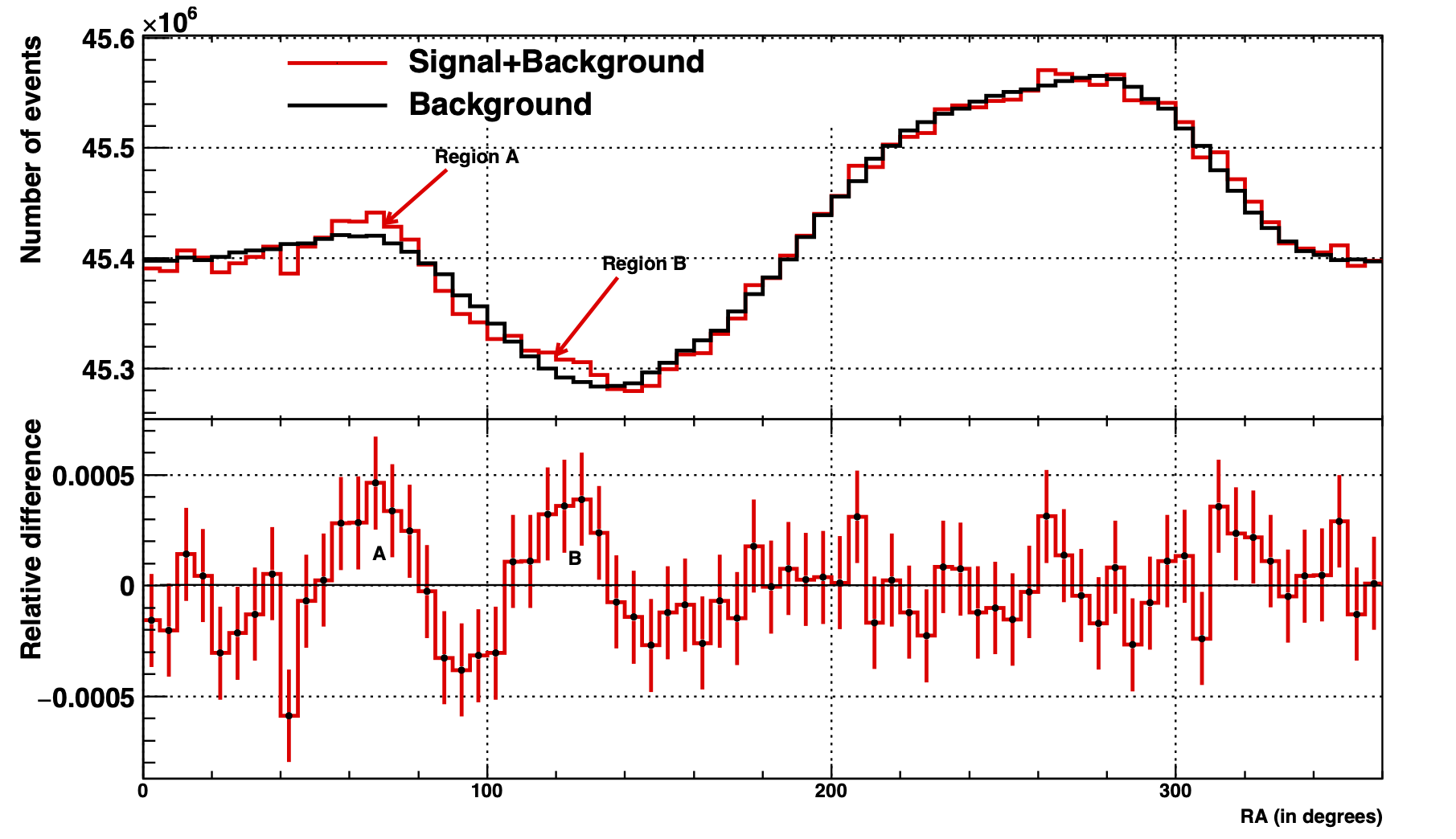}
    \caption{1-d projection of anisotropy showing the variation in number of events with right ascension}
    \label{fig:fig4}
\end{figure}

\begin{figure}[ht]
  \centering
   \subfloat{\includegraphics[width=0.8\linewidth]{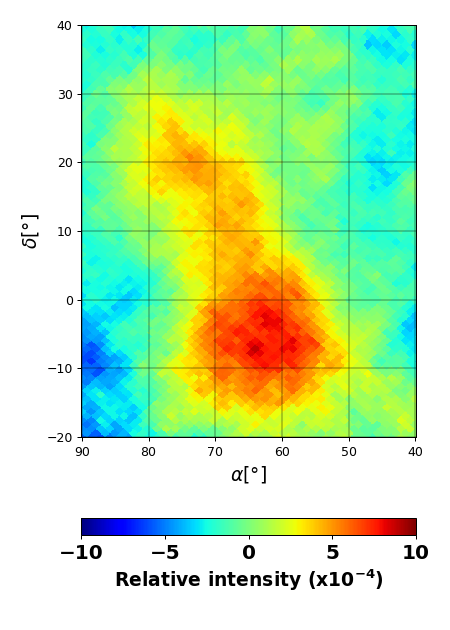}}
  \qquad
  \subfloat{\includegraphics[width=0.8\linewidth]{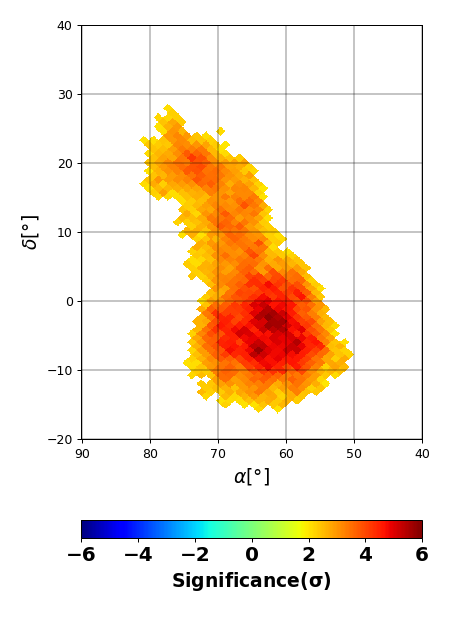}} 
 \caption{Anisotropy (top) and significance (bottom) of Region A as observed with a scrambling window of 4 hrs. The pixels with higher than 2$\sigma$ significance have been shown in the bottom plot and the region defined has been used to calculate the significance of the entire structure. }  \label{fig:fig5}
\end{figure}

\begin{figure}[ht]
\centering
  \subfloat{\includegraphics[width=0.8\linewidth]{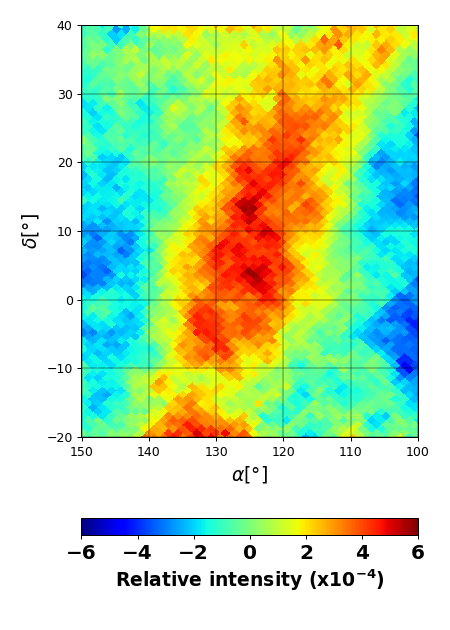}}
  \qquad
  \subfloat{\includegraphics[width=0.8\linewidth]{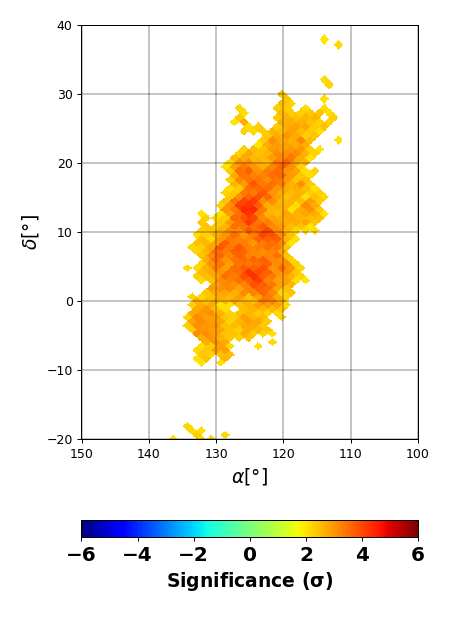}} 
 \caption{Anisotropy (top) and significance (bottom) of Region B as observed with a scrambling window of 4 hrs, and plotted similar to region A}
  \label{fig:fig6}
\end{figure}

The variation of events as a function of right ascension, obtained from the unsmoothed maps by summing events over declination bands spanning within $-18.6^{\circ}$ to $41.4^{\circ}$, is shown in \autoref{fig:fig4}. The range is taken within $\pm$ 30$^{\circ}$ from the latitude of the GRAPES-3 experiment. The plot clearly illustrates that the variation in the event count, resulting from non-uniform sky exposure due to detector and atmospheric effects, is mirrored in the background estimation obtained through the time-scrambling algorithm. The systematic excesses for regions A and B are also seen in the relative intensity plot. A zoomed view of the two regions are shown in \autoref{fig:fig5} and \autoref{fig:fig6}. The details are discussed below.  

\begin{figure}[ht]
\centering
   \subfloat{\includegraphics[scale=0.33]{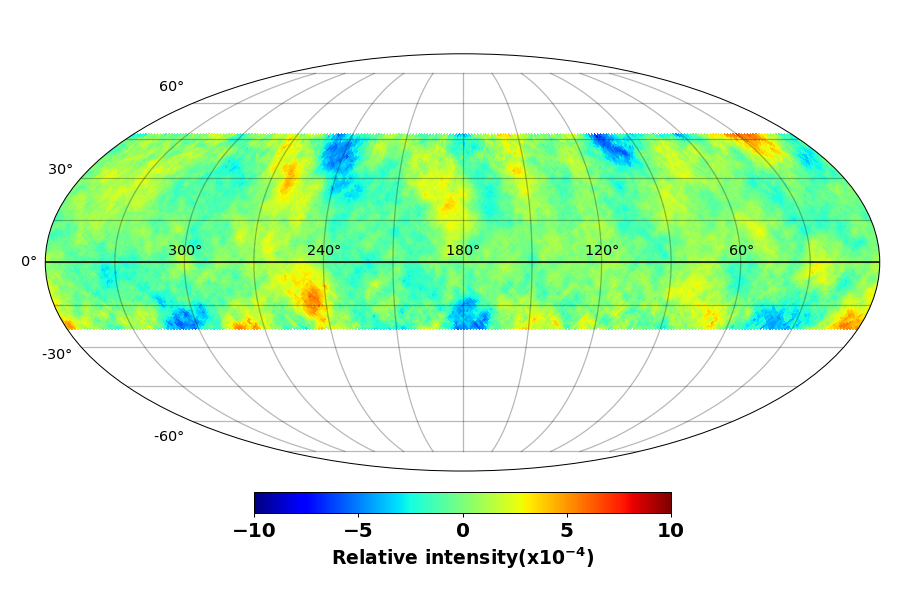}}
  \qquad
  \subfloat{\includegraphics[scale=0.33]{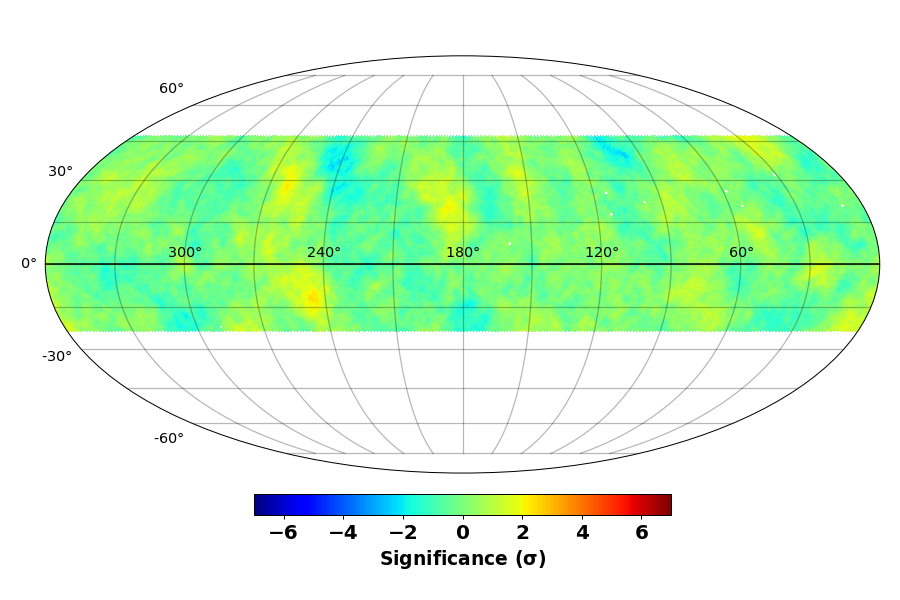}} 
 \caption{Anisotropy (top) and significance (bottom) obtained after performing the same analysis with anti-sidereal time. No characteristic signal region can be observed.}
  \label{fig:fig7}
\end{figure}

Region A is observed in the right ascension range of $\sim$50$^{\circ}$ to $80^{\circ}$ and declination range of $\sim$-15$^{\circ}$ to 30$^{\circ}$. The maximum relative intensity of the anisotropy in this region is $(8.9\pm 2.1 \pm 0.2)\times 10^{-4}$ at the pixel centered at ($\alpha=63.9^{\circ}$, $\delta=-7.2^{\circ}$), where the first error is statistical and the second error is systematic. The calculation of the systematic error involves conducting an analysis using ``anti-sidereal" time \citep{Nagashima1998, Abdo2008}. While sidereal time corresponds to the sky fixed frame, anti-sidereal time represents a non-physical frame. Analyzing data based on sidereal time allows us to determine the presence of anisotropy, while analyzing data based on anti-sidereal time provides insight into the systematic error on sidereal anisotropy resulting from non-physical effects \citep{Abbasi2010, Farley1954}. The same time-scrambling analysis with a time window of 4 hrs  was performed but with anti-sidereal time in order to estimate the systematics and the maximum strength observed in regions A and B are quoted as systematic errors. The results of anti-sidereal time have been shown in \autoref{fig:fig7} and no characteristic signal regions can be seen in this case implying that the systematics caused by non-physical effects are insignificant. The statistical error dominates as the  systematic error is much lesser than statistical error.

Region A appears to be a circular structure with a tail like projection. The maximum pixel significance observed in this region is 5.8$\sigma$. To calculate the total significance of this region, the unsmoothed data and reference maps were used in order to avoid the correlations introduced by smoothing between the pixels. First, the structure was defined by selecting those pixels which have a significance of more than 2$\sigma$ as shown in \autoref{fig:fig5} in bottom. The total number of events in data and reference maps from the region was obtained by summing up the pixel wise events in the selected area. The Li-Ma prescription was used to obtain the total significance which is 6.8$\sigma$. The relative excess number of events in this region is $(6.5\pm 1.3) \times 10^{-4}$. 

Region B is an elongated structure observed within $\sim$110$^{\circ}$ to 140$^{\circ}$ of right ascension and almost throughout the full declination range (see \autoref{fig:fig6}). The maximum relative intensity observed for this region is $(5.6\pm 1.8 \pm 0.1 )\times10^{-4}$ at the pixel centered at ($\alpha=124.5^{\circ}, \delta=3.4^{\circ}$) and the significance of the pixel is 4.4$\sigma$. Similar to the criteria applied for region A, those pixels which have a significance of more than 2$\sigma$ were selected to define region B. The overall relative intensity and significance of the region B is $(4.9\pm 1.4)\times 10^{-4}$ and 4.7$\sigma$, respectively. 

The deficit seen around regions A and B are also consistent with the observations by Milagro, ARGO-YBJ and HAWC. The deficit observed between regions A and B is the most significant. The deficit structure has a significance of $3.7\sigma$ and a relative intensity of $-(4.6\pm 1.7)\times10^{-4}$.

\section{Discussion} \label{sec:discus}
By analyzing 3.7$\times$10$^{9}$ EAS events collected over a period of 4 years, GRAPES-3 could significantly observe two excess regions namely A and B. The region A shows a tail like structure in the Northern hemisphere ($\delta>0^{\circ}$). The shape of the structure is similar to the ``region A" observed by Milagro \citep{Abdo2008} and HAWC \citep{Abeysekara2014}, and ``region 1" reported by ARGO-YBJ \citep{Bartoli2013}. Milagro (at $36^{\circ}$N) observes the part of this structure in the Northern sky and the observations are continued to the Southern hemisphere by ARGO-YBJ (at $30^{\circ}$N), HAWC (at $19^{\circ}$N) and GRAPES-3 (at $11.4^{\circ}$N). GRAPES-3 and HAWC lying closer to the Equator have the advantage of covering the southern part of region A. Region B is also observed by Milagro, ARGO-YBJ (referred to as ``region 2") and HAWC as a continuous structure running almost throughout the entire declination band, similar to observations by GRAPES-3. The full sky analysis by HAWC and IceCube show that region B continues to the Southern hemisphere as well \citep{Abeysekara2019} running through the entire declination band. The deficit regions seen around these excesses by GRAPES-3 are also coincident with observations by Milagro, ARGO-YBJ and HAWC.

The highest observed peak relative intensities for region A are $(8.5\pm 0.6 \pm 0.8)\times 10^{-4}$ as measured by HAWC \citep{Abeysekara2014}, $10.0 \times 10^{-4}$ by ARGO-YBJ \citep{Bartoli2013} and $(8.9 \pm 2.1 \pm 0.3)\times 10^{-4}$ by GRAPES-3. Region A's peak intensity is situated in the Southern hemisphere at $\delta = -7.2^{\circ}$ and $-6.3^{\circ}$ for GRAPES-3 and HAWC respectively.  Region B exhibits a peak relative intensity of $(5.2 \pm 0.6 \pm 0.7) \times 10^{-4}$ for HAWC, $5.0\times10^{-4}$ for ARGO-YBJ and $(5.6\pm1.8\pm0.1)\times10^{-4}$ for GRAPES-3. The peak intensities have been tabulated in Table \ref{tab:Table1}.
\begin{table}
\centering
\begin{tabular}{|c|c|c|}\hline
 & Region A ($\times10^{-4}$) & Region B ($\times10^{-4}$) \\ \hline
 ARGO-YBJ & 10.0 & 5.0\\
 HAWC     & $(8.5\pm 0.6\pm 0.8)$ & $(5.2\pm 0.6\pm 0.7)$ \\
 GRAPES-3 & $(8.9\pm 2.1 \pm 0.3)$ & $(5.6\pm 1.8\pm 0.1)$\\ \hline

     \hline
     
\end{tabular}
\caption{The peak intensities of regions A and B as reported by ARGO-YBJ, HAWC and GRAPES-3}
\label{tab:Table1}
\end{table}

ARGO-YBJ \citep{Bartoli2013} and HAWC \citep{Abeysekara2014} have also performed an analysis based on energy dependence by partitioning their datasets into multiple segments, some of which overlap with the median energy range of GRAPES-3 at about 16 TeV. The relative intensity of region A observed by GRAPES-3 is $(6.5\pm1.3)\times10^{-4}$. ARGO-YBJ divided their dataset into five segments, with the last two segments having median energies of 7.3 TeV and 20 TeV, respectively. In this case, the relative intensity of region A exhibited a flattening around $7.0\times10^{-4}$ in the last two energy bins which encompass GRAPES-3's median energy. When considering HAWC's analysis, their dataset was divided into seven segments, and the relative intensity of region A was assessed for each segment around the peak. It was reported by HAWC that the relative intensity of region A increases with energy. When comparing our findings, we use the segment of data having a median energy of $14.7^{+28.7}_{-9.9}$ TeV, which is the closest approximation to GRAPES-3's median energy and a relative intensity of $\sim\!(22.0\pm5.0)\times10^{-4}$ was reported. For region B, the relative intensity of the observed structure is $(4.9\pm1.4)\times10^{-4}$ by GRAPES-3. The average relative intensity of region B measured by ARGO-YBJ is $3.5\times10^{-4}$. It varies in the range of ($3.0-5.0)\times10^{-4}$ for the last two energy bins that cover GRAPES-3's median energy.

There are different explanations for the origin of the small-scale anisotropic structures. Several models propose that the origin of the hotspots might be linked to the proximity of supernova explosions, events known for generating cosmic rays (\cite{Erlykin2006}). According to (\cite{Salvati2008}), regions A and B could result from a phenomenon associated with a supernova explosion that gave rise to the Geminga pulsar. Situated between the observed structures of regions A and B is the Geminga pulsar ($\alpha,\delta$=$98.47^{\circ},17.77^{\circ}$). Some models suggest that these structures are a consequence of the turbulent magnetic field within the cosmic ray scattering length \citep{Giacinti2012, Ahlers2015} or CR scattering by Alfven waves created by turbulent cascades in local field direction \citep{Malkov2010}. Another set of models suggest that they are generated due to magnetic reconnections in the heilosphere \citep{LazarianDesiati2012}.  The model presented in \cite{Drury2008b} investigates the hypothesis that region A could potentially be explained by the production of secondary neutrons in the concentrated tail of interstellar material that forms downstream of the Sun's movement through the local interstellar medium (ISM). Some exotic models also suggest that decay of quark matter in pulsars or the acceleration of strangelets near molecular clouds are causes for the anisotropy \citep{Kotera2013, ngelesPrezGarca2014}. Thus, the examination of CR anisotropy assumes importance in comprehending the magnetic field structure, propagation, acceleration mechanisms of CRs, and their sources.

\section{Summary} 
We have investigated small-scale anisotropy in CR distribution by analyzing $3.7\times10^{9}$ events at median energy of $\sim16$ TeV, collected by the GRAPES-3 experiment from 1 January 2013 to 31 December 2016 with a live time of 1273.1 days. Time scrambling method was employed to estimate the background map. Two small-scale structures namely regions A and B have shown excesses in the CR flux at the level of ($6.5\pm1.3)\times10^{-4}$ and ($4.9\pm1.4)\times10^{-4}$, respectively with statistical significance of $6.8\sigma$ and $4.7\sigma$, respectively. These structures are in agreement with previous results reported by Milagro, ARGO-YBJ, and HAWC.

\begin{acknowledgments}

We thank D.B. Arjunan, V. Jeyakumar, S. Kingston, K. Manjunath, S. Murugapandian, S. Pandurangan, B. Rajesh, V. Santhoshkumar, M.S. Shareef, C. Shobana, R. Sureshkumar for their role in efficient running of the experiment. We would also like to thank Dr. Suvodip Mukherjee from the Department of Astronomy and Astrophysics in TIFR for his help related to parts of this work. We would also like to thank Dr. Frank McNully from Mercer University for discussing concepts related to cosmic ray anisotropy. We acknowledge the support of the Department of Atomic Energy, Government of India, under Project Identification No. RTI4002. This work was partially supported by grants from Chubu University, Japan. 

\end{acknowledgments}

\bibliography{references}{}

\begin{thebibliography}{}
\expandafter\ifx\csname natexlab\endcsname\relax\def\natexlab#1{#1}\fi
\providecommand{\url}[1]{\href{#1}{#1}}
\providecommand{\dodoi}[1]{doi:~\href{http://doi.org/#1}{\nolinkurl{#1}}}
\providecommand{\doeprint}[1]{\href{http://ascl.net/#1}{\nolinkurl{http://ascl.net/#1}}}
\providecommand{\doarXiv}[1]{\href{https://arxiv.org/abs/#1}{\nolinkurl{https://arxiv.org/abs/#1}}}

\bibitem[{{A.~A. Abdo} {et~al.}(2008)}]{Abdo2008}
{A.~A. Abdo}, {et~al.} 2008, Phys. Rev. Lett., 101,
  \dodoi{10.1103/physrevlett.101.221101}

\bibitem[{{A.~A. Abdo} {et~al.}(2009)}]{Abdo2009}
---. 2009, The Astrophys. J., 698, 2121, \dodoi{10.1088/0004-637X/698/2/2121}

\bibitem[{{A. Aab} {et~al.}(2018)}]{Aab2018}
{A. Aab}, {et~al.} 2018, The Astrophys. J., 868, 4,
  \dodoi{10.3847/1538-4357/aae689}

\bibitem[{{A. U. Abeysekara} {et~al.}(2014)}]{Abeysekara2014}
{A. U. Abeysekara}, {et~al.} 2014, The Astrophys. J., 796, 108,
  \dodoi{10.1088/0004-637x/796/2/108}

\bibitem[{{A. U. Abeysekara} {et~al.}(2018)}]{Abeysekara2018}
---. 2018, The Astrophys. J., 865, 57, \dodoi{10.3847/1538-4357/aad90c}

\bibitem[{{A.~U. Abeysekara} {et~al.}(2019)}]{Abeysekara2019}
{A.~U. Abeysekara}, {et~al.} 2019, The Astrophys. J., 871, 96,
  \dodoi{10.3847/1538-4357/aaf5cc}

\bibitem[{{A.D. Erlykin and A.W. Wolfendale}(2006)}]{Erlykin2006}
{A.D. Erlykin and A.W. Wolfendale}. 2006, Astropart. Phys., 25, 183,
  \dodoi{10.1016/j.astropartphys.2006.01.003}

\bibitem[{{B. Bartoli} {et~al.}(2013)}]{Bartoli2013}
{B. Bartoli}, {et~al.} 2013, Phys. Rev. D, 88,
  \dodoi{10.1103/physrevd.88.082001}

\bibitem[{{B. Bartoli} {et~al.}(2018)}]{Bartoli2018}
---. 2018, The Astrophys. J., 861, 93, \dodoi{10.3847/1538-4357/aac6cc}

\bibitem[{{D.~E. Alexandreas} {et~al.}(1993)}]{Alexandreas1993}
{D.~E. Alexandreas}, {et~al.} 1993, Nucl. Instrum. Meth. A, 328, 570,
  \dodoi{10.1016/0168-9002(93)90677-A}

\bibitem[{{D. Pattanaik} {et~al.}(2022)}]{Pattanaik2022}
{D. Pattanaik}, {et~al.} 2022, Phys. Rev. D, 106,
  \dodoi{10.1103/physrevd.106.022009}

\bibitem[{{F. J. M. Farley and J. R. Storey}(1954)}]{Farley1954}
{F. J. M. Farley and J. R. Storey}. 1954, Proceedings of the Physical Society.
  Section A, 67, 996, \dodoi{10.1088/0370-1298/67/11/306}

\bibitem[{{{G. Giacinti and G. Sigl}}(2012)}]{Giacinti2012}
{{G. Giacinti and G. Sigl}}. 2012, Phys. Rev. Lett., 109, 071101,
  \dodoi{https://doi.org/10.1103/PhysRevLett.109.071101}

\bibitem[{{K. Greisen}(1960)}]{Greisen1960}
{K. Greisen}. 1960, Ann. Rev. of Nucl. Sci., 10, 63,
  \dodoi{10.1146/annurev.ns.10.120160.000431}

\bibitem[{{K. Kamata and J. Nishimura}(1958)}]{kamata1958}
{K. Kamata and J. Nishimura}. 1958, Prog. of Theo. Phys. Suppl., 6, 93,
  \dodoi{10.1143/ptps.6.93}

\bibitem[{{K. Kotera} {et~al.}(2013)}]{Kotera2013}
{K. Kotera}, {et~al.} 2013, Phys. Lett. B, 725, 196,
  \dodoi{10.1016/j.physletb.2013.07.010}

\bibitem[{{K. M. Gorski} {et~al.}(2005)}]{Gorski2005}
{K. M. Gorski}, {et~al.} 2005, The Astrophys. J., 622, 759,
  \dodoi{10.1086/427976}

\bibitem[{{K. Nagashima} {et~al.}(1998)}]{Nagashima1998}
{K. Nagashima}, {et~al.} 1998, Journ. of Geophys. Res.: Space Physics, 103,
  17429, \dodoi{10.1029/98ja01105}

\bibitem[{{L. Drury and F. Aharonian}(2008)}]{Drury2008b}
{L. Drury and F. Aharonian}. 2008, Astropart. Phys., 29, 420,
  \dodoi{10.1016/j.astropartphys.2008.04.007}

\bibitem[{{M. Aglietta} {et~al.}(2009)}]{Aglietta2009}
{M. Aglietta}, {et~al.} 2009, The Astrophys. J., 692, L130,
  \dodoi{10.1088/0004-637x/692/2/l130}

\bibitem[{{M. Ahlers}(2019)}]{Ahlers2019}
{M. Ahlers}. 2019, The Astrophys. J., 886, L18,
  \dodoi{10.3847/2041-8213/ab552f}

\bibitem[{{M. Ahlers} {et~al.}(2016)}]{Ahlers2016}
{M. Ahlers}, {et~al.} 2016, The Astrophys. J., 823, 10,
  \dodoi{10.3847/0004-637x/823/1/10}

\bibitem[{{M. Ahlers and P. Mertsch}(2015)}]{Ahlers2015}
{M. Ahlers and P. Mertsch}. 2015, The Astrophys. J., 815, L2,
  \dodoi{10.1088/2041-8205/815/1/l2}

\bibitem[{{M. Amenomori} {et~al.}(2006)}]{Amenomori2006}
{M. Amenomori}, {et~al.} 2006, Science, 314, 439,
  \dodoi{10.1126/science.1131702}

\bibitem[{{M. Amenomori} {et~al.}(2017)}]{Amenomori2017}
---. 2017, The Astrophys. J., 836, 153, \dodoi{10.3847/1538-4357/836/2/153}

\bibitem[{{M. {\'{A}}ngeles P{\'{e}}rez-Garc{\'{\i}}}
  {et~al.}(2014)}]{ngelesPrezGarca2014}
{M. {\'{A}}ngeles P{\'{e}}rez-Garc{\'{\i}}}, {et~al.} 2014, Nucl. Instrum. and
  Meth. in Phys. Res. Section A: Accelerators, Spectrometers, Detectors and
  Associated Equipment, 742, 237, \dodoi{10.1016/j.nima.2013.11.007}

\bibitem[{{M.~G. Aartsen} {et~al.}(2016)}]{Aartesen2016}
{M.~G. Aartsen}, {et~al.} 2016, The Astrophys. J., 826, 220,
  \dodoi{https://doi.org/10.3847/0004-637X/826/2/220}

\bibitem[{{M. Salvati and B. Sacco}(2008)}]{Salvati2008}
{M. Salvati and B. Sacco}. 2008, Astron. and Astrophys., 485, 527,
  \dodoi{10.1051/0004-6361:200809586}

\bibitem[{{M. Zuberi} {et~al.}(2017)}]{Meeran2017}
{M. Zuberi}, {et~al.} 2017, in Proceedings of 35th International Cosmic Ray
  Conference {\textemdash} {PoS}({ICRC}2017) No. 302 (Sissa Medialab),
  \dodoi{10.22323/1.301.0302}

\bibitem[{Malkov {et~al.}(2010)Malkov, Diamond, Drury, \& Sagdeev}]{Malkov2010}
Malkov, M.~A., Diamond, P.~H., Drury, L.~O., \& Sagdeev, R.~Z. 2010, The
  Astrophys. J., 721, 750, \dodoi{10.1088/0004-637x/721/1/750}

\bibitem[{{{P. Blasi and E. Amato }}(2012)}]{Blasi2012}
{{P. Blasi and E. Amato }}. 2012, J. Cosmol. Astropart. Phys., 01, 011,
  \dodoi{10.1088/1475-7516/2012/01/010}

\bibitem[{{P. Desiati and A. Lazarian}(2012)}]{LazarianDesiati2012}
{P. Desiati and A. Lazarian}. 2012, in American Inst. of Phys. Conf. Series,
  Vol. 1436, Physics of the Heliosphere: A 10 Year Retrospective, 163--170,
  \dodoi{10.1063/1.4723604}

\bibitem[{{P.~K. Mohanty} {et~al.}(2009)}]{Mohanty2009}
{P.~K. Mohanty}, {et~al.} 2009, Astropart. Phys., 31, 24,
  \dodoi{10.1016/j.astropartphys.2008.11.004}

\bibitem[{{{P. Mertsch and S. Funk}}(2015)}]{Mertsch2015}
{{P. Mertsch and S. Funk}}. 2015, Phys. Rev. Lett., 114, 021101,
  \dodoi{10.1103/PhysRevLett.114.021101}

\bibitem[{{R. Abbasi} {et~al.}(2010)}]{Abbasi2010}
{R. Abbasi}, {et~al.} 2010, The Astrophys. J., 718, L194,
  \dodoi{10.1088/2041-8205/718/2/l194}

\bibitem[{{R.~U. Abbasi} {et~al.}(2020)}]{Abbasi2020}
{R.~U. Abbasi}, {et~al.} 2020, The Astrophys. J., 898, L28,
  \dodoi{10.3847/2041-8213/aba0bc}

\bibitem[{{S.~K. Gupta} {et~al.}(2005)}]{Gupta2005}
{S.~K. Gupta}, {et~al.} 2005, Nucl. Instrum. and Meths. in Phys. Res. Section
  A: Accelerators, Spectrometers, Detectors and Associated Equipment, 540, 311,
  \dodoi{10.1016/j.nima.2004.11.025}

\bibitem[{{{T.~P. Li and Y.~Q. Ma}}(1983)}]{LiMa}
{{T.~P. Li and Y.~Q. Ma}}. 1983, The Astrophys. J., 272, 317,
  \dodoi{https://doi.org/10.1086/161295}

\bibitem[{{V.B. Jhansi} {et~al.}(2020)}]{Jhansi2020}
{V.B. Jhansi}, {et~al.} 2020, Journ. of Cosmol. and Astropart. Phys., 2020,
  024, \dodoi{10.1088/1475-7516/2020/07/024}

\bibitem[{{Y. Hayashi} {et~al.}(2005)}]{Hayashi2005}
{Y. Hayashi}, {et~al.} 2005, Nucl. Instrum. and Meth. in Phys. Res. Section A:
  Accelerators, Spectrometers, Detectors and Associated Equipment, 545, 643,
  \dodoi{10.1016/j.nima.2005.02.020}

\end{thebibliography}
\bibliographystyle{aasjournal}



\end{document}